\documentclass[11pt]{article} %layout=twocolumn
\usepackage[version=3]{mhchem} % Formula subscripts using \ce{}
\usepackage[usenames, dvipsnames]{color}%Just for editing manuscript
\usepackage{graphicx}
\author{E Michoulier
\thanks{Univ. Lille, CNRS, UMR 8523 -- PhLAM -- Laboratoire de Physique des Lasers Atomes et Mol\'{e}cules, F-59000 Lille, France. \& Laboratoire de Chimie et Physique Quantiques LCPQ/IRSAMC, Universit\'e de Toulouse and CNRS, UT3-Paul Sabatier,  118 Route de Narbonne, F-31062 Toulouse, France.}; C Toubin$\rm{^*}$; A Simon$\rm{^*}$; J Mascetti\thanks{Univ. Bordeaux, CNRS, Bordeaux INP, Institut des Sciences Mol\'{e}culaires UMR 5255, F-33405 Talence, France.}; C Aupetit$\rm{^\dag}$; J A Noble\thanks{CNRS, Aix Marseille Univ., Laboratoire Physique des Interactions Ioniques et Mol\'{e}culaires (PIIM, UMR 7345), Marseille, France. jennifer.noble@univ-amu.fr}}

\title{Perturbation of the Surface of Amorphous Solid Water by the Adsorption of Polycyclic Aromatic Hydrocarbons}

\begin{document}
\maketitle
%%%%%%%%%%%%%%%%%%%%%%%%%%%%%%%%%%%%%%%%%%%%%%%%%%%%%%%%%%%%%%%%%%%%%

%\textcolor{red}{Letters are limited to 2500 words or the equivalent (8--10 double-spaced typewritten pages of text, 3--4 figures, and 1--2 schemes/illustrations). A brief abstract of no more than 150 words should be included. Letters must contain a Table of Contents (TOC)/Abstract graphic as part of the manuscript.}

%%%%%%%%%%%%%%%%%%%%%%%%%%%%%%%%%%%%%%%%%%%%%%%%%%%%%%%%%%%%%%%%%%%%%

\begin{abstract}

This joint theoretical and experimental study establishes that the adsorption of polycyclic aromatic hydrocarbons (PAHs) onto the amorphous ice surface provokes a broadening and redshift of the ``dangling'' OH (dOH) ice spectral feature, the redshift increasing with PAH size up to $\sim$ 85~cm$^{-1}$. It also reveals that, in certain interaction configurations, adsorption induces substantial reorganisation of the hydrogen-bonding network at the ice surface.  Comparison with experiments validates the novel theoretical methodology relying on the density functional based tight binding approach, which offers a compromise between system size and accuracy enabling a wide sampling of surface structures.
Applied in an astrophysical context, this study suggests that widening of the dOH feature by adsorption of aromatic molecules could explain its absence heretofore in observational ice spectra, offering hope that future missions with higher sensitivity will verify its presence or absence in dense regions.

\end{abstract}

\begin{figure} %The surrounding frame is 9\,cm by 3.5\,cm
\centering
\includegraphics[width=9cm]{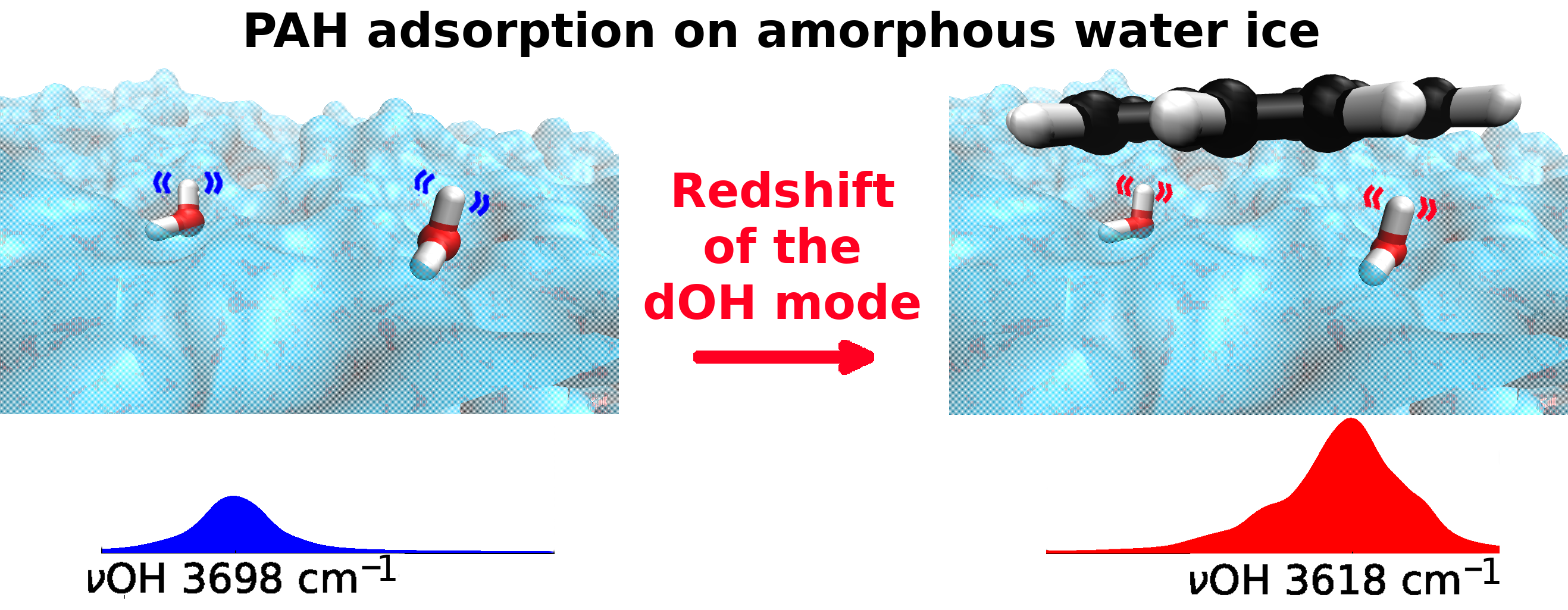} 
\caption{Table of Contents graphic}
\end{figure}

\clearpage
%\scriptsize
%%---------------------
%%SECTION: INTRODUCTION
%%---------------------
\section{Introduction}

One of the most characteristic features of the mid infrared (MIR) spectrum of amorphous solid water (ASW) is the presence of two weak absorption bands at $\sim$~3700~cm$^{-1}$, attributed to free OH bonds at the ice surface and commonly referred to as ``dangling'' OH bands (dOH).\cite{Buch91,Rowland91,Rowland95,Devlin95,Noble14a}
Although the spectral signatures of water ice in dense molecular clouds and cores point to an amorphous structure,\cite{Boogert15} these dOH bands have, to date, not been observed in interstellar ices.
Only one potential observation of a band in the dOH region has been made; a weak feature at 3660~cm$^{-1}$ was detected along the line of sight toward protostellar object S140:IRS1, and tentatively assigned to the CO$_2$ molecule adsorbed on dOH.\cite{Keane01}
However, the absence of a positive detection does not necessarily preclude the presence of these features, as the instruments used to date have not had the combination of sensitivity and spectral resolution necessary to reveal the weak dOH bands.
The lack of definitive observational evidence has given rise to debate surrounding the porosity of the amorphous water formed on dust grains, with uncertainty remaining as to the exact role formation\cite{Dulieu10} and processing\cite{Palumbo06,Accolla11,Noble14a,Noble14b,Dartois15,Mejia15} mechanisms play in determining the ice structure, and in particular its porosity.
Laboratory studies have revealed, amongst other things, that the dOH is redshifted by the adsorption of molecules onto the ice surface.\cite{Rowland95}
Small molecules present in icy interstellar environments (i.e. clouds, cores, and disks) have been shown to induce relatively small shifts on the order of a few tens of wavenumbers. The redshift induced by the adsorption of CH$_4$ on an ASW surface is on the order of 20~cm$^{-1}$, increasing to $\sim$~40~cm$^{-1}$ for CO$_2$, and 60~cm$^{-1}$ for CO.\cite{Rowland95,Palumbo06,Martin02,Manca01A}

In the interstellar medium, PAHs are predicted to represent a significant fraction of the C budget, and are generally believed to contribute to the widely observed set of MIR spectral emission features known as the ``aromatic infrared bands'' (AIBs).\cite{Tielens13} They have also been proposed as the carriers of some of the ``diffuse interstellar bands'' (DIBs), a series of absorption bands in the near-IR--visible domain observed on the extinction curve of our galaxy.\cite{herbig95,Salama11}
If PAHs are present in the interstellar medium, they are expected to freeze out onto dust grains in molecular clouds and form part of their icy mantles, within which they may be implicated in reaction networks. Once removed from the gas phase, their IR emission would be attenuated and they would only be observable in absorption, typically contributing to the complex 5--8~$\mu$m region.\cite{Boogert08} Although no such IR absorption feature has yet been unambiguously attributed to PAHs, it is clear that the interaction of PAHs with ice is key in determining both their physical chemistry and their observational signatures in such cold, dense environments.
Another signature which could be indicative of molecular adsorption onto icy mantles is the dOH.
Several experimental studies on individual PAH molecules have shown that the dOH redshift induced by PAH adsorption is significantly larger than that of small molecules, with reported values up to $\sim$~100-110~cm$^{-1}$.\cite{Silva94,Guennoun11c,Guennoun11p}

To our knowledge, there has been no prior systematic study (experimental or theoretical) of the interaction of PAHs of increasing sizes with water ice and, critically, their interaction under interstellar-relevant conditions has not been modelled in terms of the influence on the dOH feature. From the theoretical point of view, the complexity hinges upon modelling a realistic PAH-ice system with both an explicit description of the electronic structure -- mandatory in accounting for the vibrational structure -- and a statistical description of the possible adsorption geometries including an explicit water environment. So far, this has not been possible due to the large sizes of the systems to be considered and the associated computational cost.
In this work, we present the first joint experimental and computational study describing the influence of four aromatic molecules of increasing size (one to six rings) on the dOH feature. We present the experimental data before introducing a robust, newly-developed theoretical methodology, building on our previous studies of PAH-ice interactions\cite{Michoulier18a,Michoulier18b} to accurately describe the influence of the adsorption of molecules at an ice surface on the dOH vibrational features.

%\vspace{5mm}

\section{Experimental and theoretical methods}
\subsection{Experimental methods}

Porous amorphous solid water (pASW) samples were prepared by depositing molecules onto a CsBr window held at 15~K in a high vacuum chamber. The setup has been detailed previously.\cite{Simon17,Guennoun11p} Molecules were introduced by vapour deposition after multiple freeze-pump-thaw cycles (in the case of water and benzene (Bz, C$_6$H$_6$)) or by heating an oven positioned in front of the cold surface (anthracene (Anth, C$_{14}$H$_{10}$), pyrene (Pyr, C$_{16}$H$_{10}$), and coronene (Cor, C$_{24}$H$_{12}$)). FTIR spectra of 200 scans were obtained in transmission mode using a Bruker 70~V spectrometer. Co-deposition of water and aromatic molecules maximises the interactions between aromatics and the pASW surface, providing optimal IR signal for studying the surface water modes. The relative concentrations of water and the aromatics were also chosen to optimise interactions, with the values, determined via analysis of the vibrational bands, being 0.5--2~\% for the PAHs\cite{Bouwman11,Cook15,deBarros17} and $\sim$~50~\% for Bz.\cite{Yamada83} An additional Bz:H$_2$O sample of $\sim$~3~\% is included for completeness.

\subsection{Theoretical methods}

The modelling of PAH-ice systems undertaken in this work is based on previous studies.\cite{Michoulier18a,Michoulier18b}
Theoretical descriptions of aromatic-ice systems (Bz, Anth, Pyr, Cor) were based on starting point geometries extracted from classical molecular dynamics (MD) trajectories for Low Density Amorphous (LDA) ices.\cite{Michoulier18a} The electronic structure is explicitly described for all atoms at the Self Consistent Charge Density Functional based Tight Binding (SCC-DFTB) level \cite{Elstner98} with a modified version of the Hamiltonian dedicated to the study of molecular clusters, for which the description of long range interactions is improved \textit{i.e.} with CM3 charges \cite{Li_CM3} replacing Mulliken charges and empirical dispersion corrections.\cite{DFTB_CM3,joalland2010} This method, designated hereafter as ``DFTB'' for simplicity, has been extensively benchmarked and used to describe benzene and PAHs in interaction with water clusters.\cite{SimonPCCP2012,SimonJCP2013,Simon19} 
Additional benchmarking, presented below, was performed on benzene (Bz) interacting with (H$_2$O)$_8$ clusters.

In this work, the modelled aromatic-ice systems are four sets of 49 finite-sized local minima consisting of one aromatic molecule in interaction with an ice cluster of an average size of 50 central water molecules that are allowed to relax and 90 frozen peripheral molecules (to maintain ice structure).\cite{Michoulier18b} The local minima previously obtained using ``weighted Mulliken charges'' (WMULL)\cite{Michoulier18b} were re-optimized at the DFTB level, \textit{i.e.} using CM3 charges, which our benchmarking on Bz(H$_2$O)$_8$ clusters revealed to give more satisfactory results than WMULL in terms of harmonic vibrations, although both approaches are quite similar.

The harmonic IR spectra of both pure ice and aromatic-ice systems were obtained by diagonalisation of the weighted Hessian matrix, in which the off-diagonal terms coupling the mobile and fixed molecules were set to zero.
In the past, we have taken into consideration finite temperature anharmonic effects in the study of water clusters.\cite{SimonPCCP2012} However, in this case, the time necessary for convergence of the dynamic IR spectra for aromatic-ice systems of $\sim$ 140 molecules is prohibitively long.
In fact, the inherent width of the IR bands in the solid phase, as well as the sample averaging over tens of configurations, precludes the need to take into consideration the minor broadening effect due to anharmonicity, while the application of a scaling factor accurately reproduces shifts.

\subsection{DFTB benchmarking for dOH mode and weak interactions}

Benchmarking was performed on Bz interacting with (H$_2$O)$_8$ clusters, \textit{i.e.} Bz(H$_2$O)$_8$, focusing on the description of the dOH feature and of its interaction with an aromatic molecule. This model system was chosen due to the availability of gas phase experimental IR spectra.\cite{Gruenloh97,Gruenloh98} The DFTB Hamiltonian used to describe these systems was modified as follows:  empirical dispersion and CM3 charges (with dOH, dCH, and dOC values of 0.12873, 0.098, and 0.0, respectively, as determined by Simon et al.\cite{SimonPCCP2012}) were added to improve the description of intermolecular interactions. 

%%Figure 1
\begin{figure*}[htb!]
\includegraphics[width=0.5\textwidth]{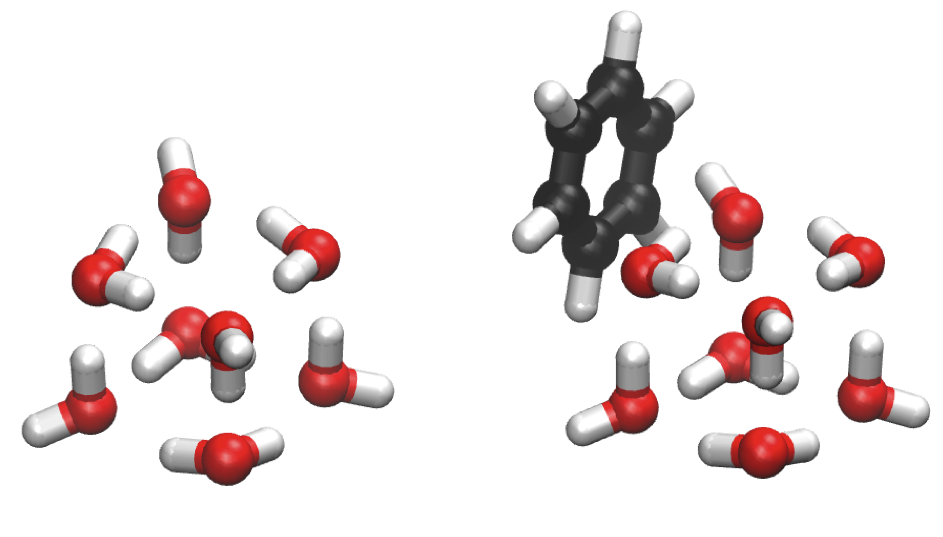}
\caption{DFTB optimised geometries of (H$_2$O)$_8$ (left) and Bz(H$_2$O)$_8$ (right) used for benchmarking. (H$_2$O)$_8$ belongs to the D$_{2d}$ symmetry point group.  Both clusters  correspond to local minima at the SCC-DFTB level with CM3 charges. The starting point geometries, taken from Miliordos et al.\cite{Miliordos16}  correspond to global minima at the MP2 level.}
\label{fig:supp_D2dbzW8}
\end{figure*}

The most stable form of (H$_2$O)$_8$ is known to be a cube, with the two most stable conformers belonging to the S$_4$ and D$_{2d}$ symmetry point groups. The geometries of the D$_{2d}$ isomer of (H$_2$O)$_8$, both isolated and adsorbed on Bz, were optimised at the DFTB level. The D$_{2d}$ geometry of Bz(H$_2$O)$_8$ had previously been determined as the most stable form by Miliordos et al.\cite{Miliordos16} The geometry obtained here is very similar to that obtained at the MP2, CCSD(T), and wB97XD (DFT functional including dispersion) levels of theory:\cite{Miliordos16} the benzene molecule is almost parallel to one of the faces of the octamer cube, with one primary $\pi$ Bz--HO interaction and two secondary Bz--O interactions.

%%Table 1
\begin{table*}[htb!]
\caption{Wavenumbers (in cm$^{-1}$) and intensities (in km\,mol$^{-1}$) of dO and dOH DFTB harmonic vibrational modes for the clusters presented in Figure~\ref{fig:supp_D2dbzW8}.} 
\centering
\begin{tabular}{|l|l|l|l||l|l|l|}
\hline
 & \multicolumn{3}{c||}{(H$_2$O)$_8$ D$_{2d}$}                                                                                      & \multicolumn{3}{c|}{Bz-(H$_2$O)$_8$ D$_{2d}$}                                                                                   \\ \hline
& Mode & Wavenumber  & Intensity  & Mode & Wavenumber & Intensity \\ \hline

dO   & 61                        & 3806.5                                    & 870.1                                    & 97                        & 3803.3                                    & 754.6                                    \\  \hline
dO   & 62                        & 3806.5                                    & 870.2                                    & 98                        & 3815.7                                    & 733.6                                    \\  \hline
dOH  & 63                        & 3925.0                                    & 27.1                                     & 99                        & 3869.6                                    & 109.4                                    \\  \hline
dOH  & 64                        & 3925.0                                    & 26.9                                     & 100                       & 3922.8                                    & 54.7                                     \\  \hline
dOH  & 65                        & 3925.7                                    & 4.0                                      & 101                       & 3924.3                                    & 97.7                                     \\  \hline
dOH  & 66                        & 3925.5                                    & 289.6                                    & 102                       & 3925.5                                    & 90.67                                    \\ \hline
\end{tabular}
\label{tab:harmosupp_D2dbzW8}
\end{table*}

From our calculations, the Bz-(H$_2$O)$_8$ complexation energy was determined to be 4.81~kcal\,mol$^{-1}$, in good agreement with the  {\it ab initio} value of 4.91~kcal\,mol$^{-1}$ obtained at the MP2/aug-cc-pvtz level including basis set correction errors.\cite{Miliordos16} 
Regarding the vibrational spectra, the interaction with Bz leads to a symmetry decrease and thus to a splitting of the degenerate modes of (H$_2$O)$_8$. 
Interesting data to compare with literature experimental results are, firstly, the occurrence of a redshift of 55~cm$^{-1}$ between the free OH stretch and the OH in $\pi$ interaction with Bz. The experimental redshift\cite{Gruenloh97, Gruenloh98} is 63~cm$^{-1}$, \textit{i.e.} our calculations underestimate the redshift by about 15~\%, and we note that a larger redshift (88~cm$^{-1}$) was obtained at the MP2 level,\cite{Miliordos16} as expected from prior theoretical studies.\cite{Zwier96} Regarding the antisymmetric double donor oxygen OH modes, the interaction with Bz results in a split of the degenerate modes resulting in two bands, one slightly redshifted (3~cm$^{-1}$) and one blueshifted by 9~cm$^{-1}$ (see data in Table~\ref{tab:harmosupp_D2dbzW8}).

Regarding the raw values of the frequencies, the comparison of our calculated value (3925~cm$^{-1}$) with the experimental value of the free OH mode resonance in (H$_2$O)$_8$ (3713.5~cm$^{-1}$) results in a scaling factor of 0.946. This is entirely unsurprising due to the approximations inherent to the method -- harmonic vibrations, approximate Hamiltonian with pre-computed pair integrals, and the minimal valence basis sets, in particular.\cite{SimonPCCP2012}
We insist on the fact that, despite the overestimation of the energy  of the OH stretching mode at the DFTB level with respect to experimental data, a good description of PAH-water cluster and water-water intermolecular interactions is expected as they were parameterised on wavefunction {\it ab initio} calculations and benchmarked in previous studies.
\cite{SimonPCCP2012,SimonJCP2013,Oliveira2015}
Therefore, the computed values of the wavenumber shifts induced by intermolecular interactions are reliable, as verified in the present study.

%%---------------------
%%SECTION: RESULTS
%%---------------------

\section{Results and discussion}

%%Figure 2
\begin{figure}[htb!]
\includegraphics[width=0.5\textwidth]{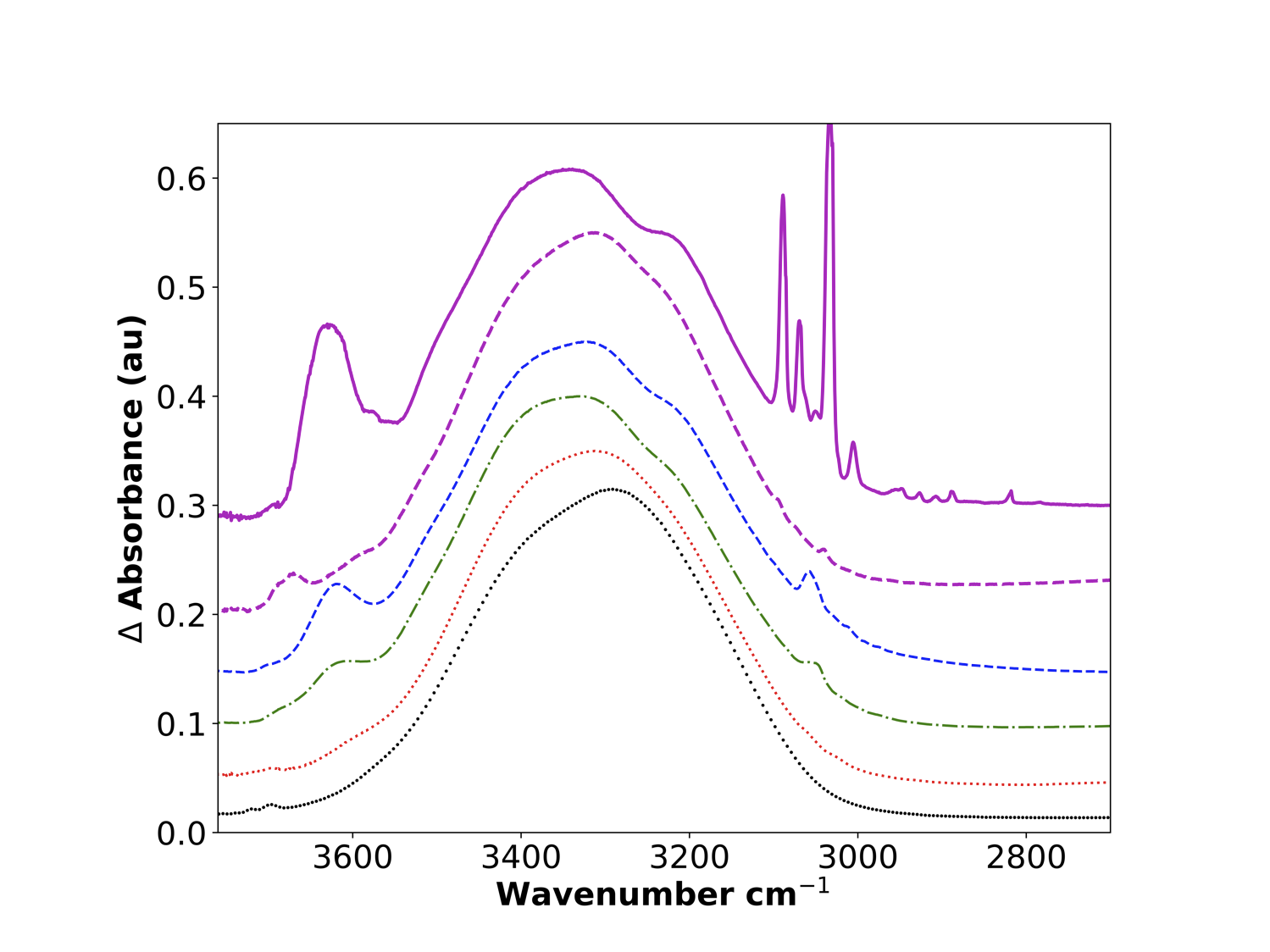}
\caption{Experimental spectra of pASW (black dots), Cor:pASW (red dotted line), Pyr:pASW (green dash-dotted line), Anth:pASW (blue dashed line), and Bz:pASW ($\sim$~3~\% dotted and $\sim$~50~\% solid magenta lines) ices deposited at 15~K. All spectra are normalised to the maximum of the OH stretching mode, and offset for clarity.} 
\label{fig:exptspectra}
\end{figure}

The experimental IR spectra of pure pASW and mixed aromatic:water ices (Bz, Pyr, Anth, Cor) are presented in Figure~\ref{fig:exptspectra}. The spectrum of pASW (black) presents a wide band at $\sim$~3300~cm$^{-1}$, corresponding to the $\nu_1$ and $\nu_3$ stretching vibrations, as well as two weak dOH features in the blue wing.\cite{Buch91,Devlin95} The addition of aromatic molecules to the ice gives rise to additional narrow features in the 3100--2800~cm$^{-1}$ region (C-H stretches, coloured spectra) as well as modification of the dOH. 
In pure pASW, the dOH features are observed at 3720 and 3699~cm$^{-1}$. 
Upon introduction of aromatic molecules, the dOH bands are redshifted, with shift values of approximately 70 (Bz, 50~\%)/ 30 (Bz, 3~\%), 78 (Anth), 79 (Pyr), and 84 (Cor)~cm$^{-1}$ compared to the strongest dOH feature at 3699~cm$^{-1}$ (Table~\ref{tab:theo}). The influence of concentration upon the Bz-ice interaction is discussed in detail below.

%\vspace{5mm}

%%Figure 3
\begin{figure*}[htb!]
\includegraphics[width=0.9\textwidth]{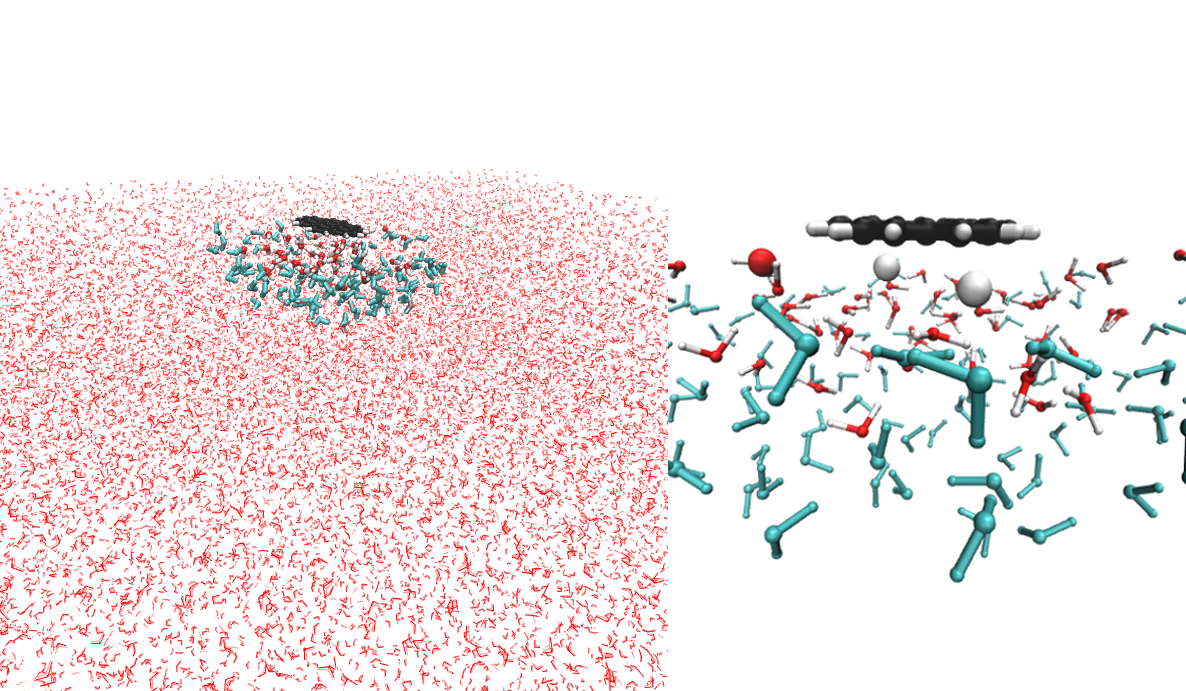}
\caption{Molecular representations of the systems modelled with the multi-method approach,\cite{Michoulier18a,Michoulier18b} illustrated by the adsorption of coronene on amorphous ice. The left hand panel presents the full structure of coronene on LDA ice at 10\,K obtained from MD/FF calculations.\cite{Michoulier18a} Molecules extracted for DFTB calculations are highlighted then magnified in the left and right panels, respectively. Water molecules in blue are fixed during DFTB optimisation and harmonic calculations, while the central 58 water molecules plus the coronene molecule remain mobile. The two white spheres are dOH in interaction with Cor. }
\label{fig:cor_lda_v2}
\end{figure*}

%Table 2
%%\begin{table}
%%\begin{tabular}{c|c|c}
%%Aromatic & $\delta$ dOH expt. & $\delta$ dOH theo.  \\
%%\hline
%%Bz    & -70 (\sim~50~\%~Bz) &  -74  \\
%%       & -30 (\sim~3~\%~Bz) & \\
%%Anth  & -78 & -79   \\
%%Pyr   & -79 & -77   \\
%%Cor   & -84 & -85   \\ 
%%\end{tabular}
%%\caption{Shift of dOH position (in cm$^{-1}$) for aromatic-ASW systems compared to pure ASW, for experimental pASW and theoretical LDA. For the theoretical values, the shifts were computed from unscaled spectra, as detailed in the text. Two experimental values are given for benzene, corresponding to the two concentrations presented as spectra in Figure~\ref{fig:exptspectra}.}\label{tab:theo}
%%\end{table}

%Table 2
\begin{table}
\begin{tabular}{c|c|c|c}
Aromatic & $\delta$dOH theo. & $\delta$dOH our expt. & $\delta$dOH literature expt. \\
\hline
Bz      &   -74 & -70 ($\sim$50~\%~Bz) &  \textit{-79 (mixed 50~\%~Bz), Dawes et al.\cite{Dawes18}} \\
          &          & -30 ($\sim$3~\%~Bz) & \textit{-111 (layered 3~\%~Bz), Silva \& Devlin.\cite{Silva94}}\\
Anth  & -79   & -78 & -- \\
Pyr    & -77    & -79 & \textit{-112 (layered), Guennoun et al.\cite{Guennoun11p}}\\
Cor   & -85    & -84 & \textit{-72 (layered), Guennoun et al.\cite{Guennoun11c}}  \\ 
\end{tabular}
\caption{Shift of dOH position (in cm$^{-1}$) for aromatic-ASW systems compared to pure ASW, for theoretical LDA and experimental pASW. For the theoretical values, the shifts were computed from unscaled spectra, as detailed in the text. Two experimental values are given for benzene, corresponding to the two concentrations studied in this work. Experimental literature data for different deposition methods is given in italics for comparison.}\label{tab:theo}
\end{table}

As explained above, the theoretical descriptions of aromatic-ice systems (Bz, Anth, Pyr, Cor) include the electronic structure explicitly at the DFTB level.
 An example of an optimised PAH-ice structure is illustrated in Figure~\ref{fig:cor_lda_v2}. A comprehensive list of calculated DFTB values for dOH positions and redshifts for pure LDA and aromatic-LDA systems is reported in Table~\ref{tab:theo}. 
The advantage of the theoretical approach presented here is its ability to compute vibrational modes for much larger systems than have been studied previously (i.e. C$_2$H$_4$, CO, O$_2$, and N$_2$\cite{Manca01A,Hujo11,Pezzella2018}), including molecules containing multiple aromatic rings, taking into consideration a sampling of geometries.

%%Table 3
\begin{table}[htb!]
\begin{tabular}{c|c|c|c}
Aromatic & pure LDA & Aromatic-LDA       & Redshift  \\\hline 
Bz    & \textit{3934} & \textit{3860} & \textit{74}\\ 
      & 3701          & 3632          & 69      \\  
Anth  & \textit{3931} & \textit{3852} & \textit{79}\\ 
      & 3699          & 3624          & 75          \\
Pyr   & \textit{3932} & \textit{3855} & \textit{77} \\
      & 3700          & 3627          & 73          \\
Cor   & \textit{3930} & \textit{3845} & \textit{85} \\
      & 3698          & 3618          & 80          \\ 
\end{tabular}
\caption{Computed positions of the dOH modes for the dangling OH in LDA alone, with an aromatic adsorbed on LDA (Aromatic-LDA), and the relative redshift values. Raw values are given in italics, with values scaled by 0.94 shown in normal type. All values are expressed in cm$^{-1}$. Positions are obtained from the mean position of the full width at half maximum. This approach yielded values very close to the average peak positions.}
\label{tab:s2}
\end{table}

\vspace{5mm}

%%Figure 4
\begin{figure}[htb!]
\includegraphics[width=0.5\textwidth]{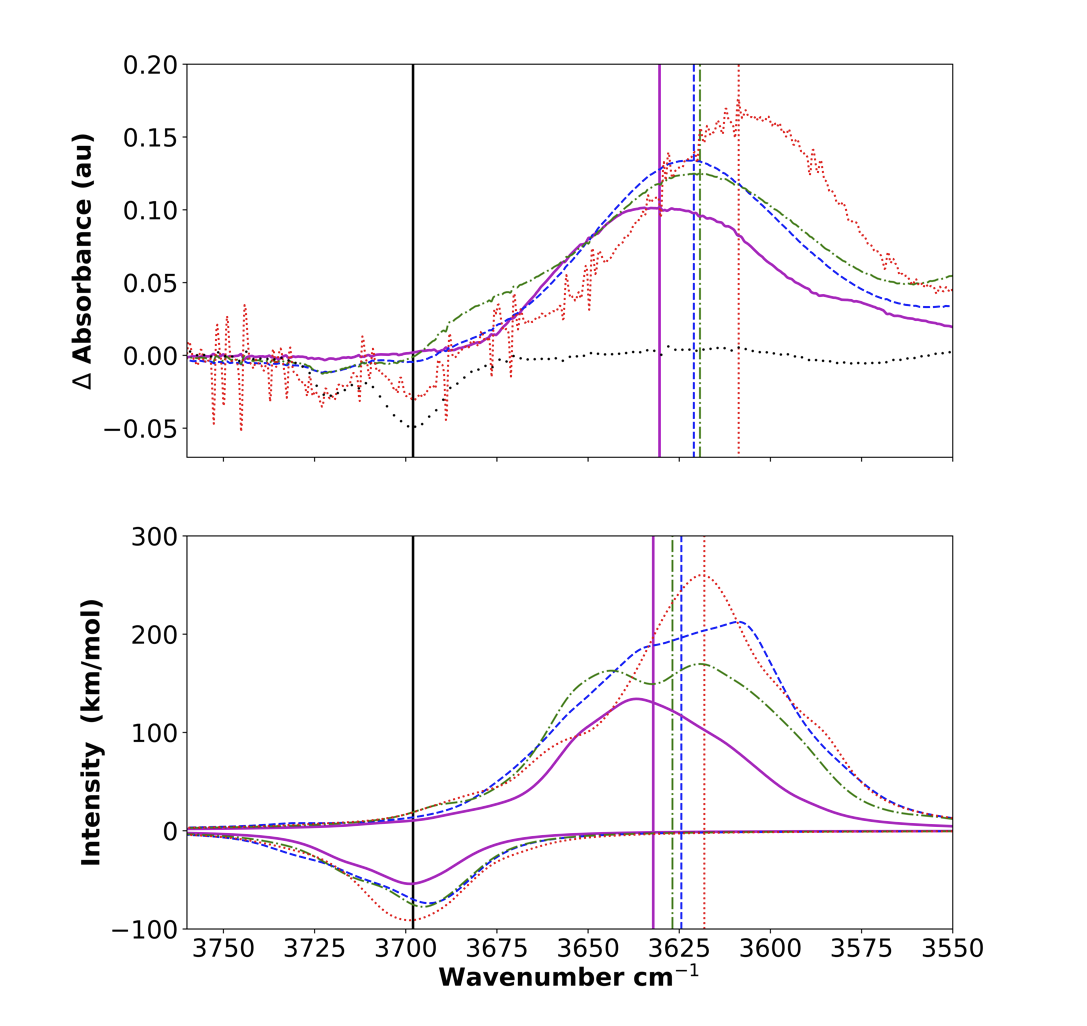}
\caption{IR spectra in the dOH region. The upper panel presents experimental pASW spectra, while the theoretical LDA spectra are presented in the lower panel. In each case, pure ASW spectra (black dotted line) are plotted in negative values to ease comparison with the aromatic-ASW spectra (Bz-ASW, solid magenta line; Anth-ASW, dashed blue line; Pyr-ASW, green dash-dotted line; Cor-ASW, dotted red line). For the sake of clarity, experimental spectra are baseline-corrected with polynomials to remove the contribution of the bulk ice OH stretching band, then scaled for plotting (factors: 2.5 Bz-pASW, 1.0 Anth-pASW, 3.0 Pyr-pASW, 15 Cor-pASW, -6.0 pASW). Similarly, the theoretical spectra were derived by selecting only dOH stretching modes from the LDA and aromatic-LDA systems, scaled by a factor of 0.94 (obtained by matching the dOH mode position in pure LDA to the experimental one at $\sim$~3699~cm$^{-1}$).}
\label{fig:compare_theoexpt}
\end{figure}

As can be observed by comparing the experimental and theoretical spectra (Figure~\ref{fig:compare_theoexpt}), the theoretical methodology employed in this work is extremely successful in modelling the aromatic-ice interaction, reproducing to within a few cm$^{-1}$ our experimental redshifts of dOH features for PAHs (see Table~\ref{tab:theo}), with the redshift value increasing with the size of the adsorbed aromatic.
It should be noted that the theoretical model also nicely replicates the adsorption-provoked broadening of the dOH band, providing confidence in the configuration sampling methodology. This effect is due to the variety of adsorption sites explored by the aromatic species at the ice surface. 
In fact, the interaction geometry is key in determining the IR signatures of binding between aromatics and water, as will be discussed below.
These results confirm the capacity of our method to describe fine vibrational perturbation induced by weak intermolecular interactions. The approach has previously proven to be highly reliable in determining the adsorption energies and ionisation potentials of aromatic molecules (from Bz to ovalene) adsorbed on ice surfaces, with the values obtained reflecting surface structural effects.\cite{Michoulier18b} However, the successful description of perturbative effects on vibrations due to weak intermolecular interactions is even more challenging than the determination of energetic data.

%%%----------
%% Discussion
%%%----------

Interestingly, when considering literature values for the redshifts observed in aromatic-ice systems, the simple mass-/size-dependent series that we determine in the current work is not necessarily evident. What is key to understanding these values is the differences in experimental methods used to produce the aromatic-ice systems. First, our result for Cor agrees relatively well with the previously published value obtained for layered ices using the same experimental setup, with the difference of $\sim$~12~cm$^{-1}$ being attributed to the differing deposition methods (layered versus mixed ices).\cite{Guennoun11c} However, our Pyr result is 33~cm$^{-1}$ blueshifted compared to a prior layered ice study.\cite{Guennoun11p} In this case, we attribute this difference to the low Pyr concentration in the previous study giving rise to uncertainty in the exact peak position of the dOH feature, and are confident that the experimental methodology applied in the present study ensures an accurate, reproducible measurement. 

The case of Bz is more complex. If we consider the two experimental Bz:pASW spectra in Figure~\ref{fig:exptspectra}, we can observe that the presence of Bz is barely seen at the lower concentration of $\sim$~3~\%, with no Bz absorption features, and a redshift in the dOH of only $\sim$~30~cm$^{-1}$. At higher concentration ($\sim$~50~\%), the presence of Bz is clearly observable via its bands in the 3100-2800~cm$^{-1}$ region, and the induced shift in the dOH is increased to $\sim$~70~cm$^{-1}$.

Of the two experimental results that we identified in the literature for Bz-ASW interactions, our value of  $\sim$~70~cm$^{-1}$ most closely resembles that of Dawes et al.\cite{Dawes18} for mixed Bz:ASW in a 1:1 concentration ($\delta_{dOH}$ 79~cm$^{-1}$). The redshift that we (and Dawes et al.) observe is, however, approximately two-thirds that obtained in the study of Silva~\&~Devlin\cite{Silva94} (111~cm$^{-1}$). In that study, the derived bond energy between benzene and the water ice surface was estimated at 4.3~kcal\,mol$^{-1}$, \textit{i.e.} twice the value between benzene and a single water molecule.\cite{Silva94} %
As for Pyr and Cor, we suspect that the source of this discrepancy is the varying deposition methods and relative concentrations applied in the different studies. 
In fact, if we also consider Bz-water cluster studies, the values of the dOH redshift can be determined to depend on both the size and geometry of the (H$_2$O)$_n$ cluster interacting with Bz. \cite{Pribble94,Zwier96,Gruenloh97,Gruenloh98,Gruenloh00} 
With respect to the highest energy free OH mode, the dOH redshift was found to vary from 61~cm$^{-1}$ (n=3) to 75~cm$^{-1}$  (n=7),\cite{Pribble94} with values of 69 and 64 cm$^{-1}$ found for  cubic-like structures  (n=8,9).\cite{Gruenloh97,Gruenloh98,Gruenloh00}

It has been demonstrated\cite{Dawes18} that varying the concentration of Bz in H$_2$O varies its interaction with the water ice surface and thus the redshifts observed. In our experiments, we observe a small shift ($\sim$~30~cm$^{-1}$) for a concentration of a few percent \textit{i.e.} an equivalent concentration to the PAHs under consideration. This suggests that Bz is not fully interacting with the surface dOH at this low concentration, as this shift is of the same order as that observed for atoms or small (diatomic) molecules adsorbed on ASW.\cite{Devlin95} Perhaps at this low concentration all or most of the Bz molecules penetrate the ice and thus are not in interaction with the ice surface in pores. When we increase the concentration, we observe an increased shift in dOH to $\sim$~70~cm$^{-1}$, suggesting that Bz is now interacting directly with the ice surface. It should be noted that at a ratio of 1:1, Bz must be interacting with water clusters, rather than being fully solvated, and thus this is more comparable to a single Bz adsorbing on an ASW surface, as in our calculations. 

The variation in the dOH redshifts in Bz-(H$_2$O)$_n$ and Bz-ice interactions highlights the primordial importance of local ice structure in the study of such phenomena, in agreement with a similar discussion in Dawes et al.\cite{Dawes18} for Bz.
This, in turn, influences the physical chemistry of the aromatics. For example, we have recently shown experimentally that the geometry of interaction governs the photoreactivity of Cor in water complexes.\cite{Noble17coro} Despite these factors, the size-dependent redshift augmentation observed in both our experimental and calculated spectra holds, whatever the Bz concentration.

\vspace{5mm}

\begin{figure}[htb!]
\includegraphics[width=0.5\textwidth]{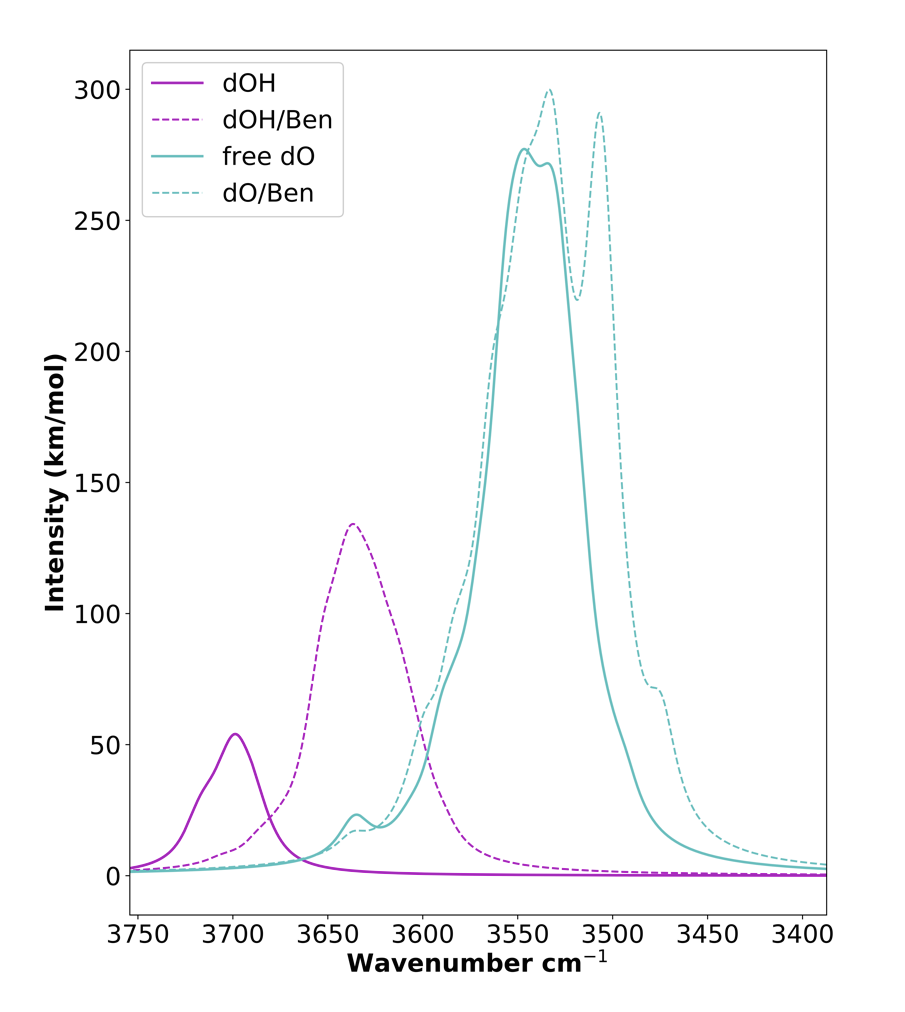}
\caption{Computed IR bands of the dO (blue) and dOH (violet) modes of LDA ice before (continuous line) and after (dashed line) adsorption of one Bz molecule (averaged over all configurations). The raw DFTB wavenumbers are scaled by 0.94. Band positions are obtained from the mean position at full width half maximum.  In pure LDA, the dOH and dO  band positions are separated by 150~cm$^{-1}$ (when scaling 0.94 is applied). Upon Bz absorption, the dOH and dO bands are redshifted respectively by 74 (69 with 0.94 factor) cm$^{-1}$ and 11 (10 with 0.94 factor) cm$^{-1}$.} 
\label{fig:ben_lda_do}
\end{figure}

\begin{figure}[htb!]
\includegraphics[width=0.5\textwidth]{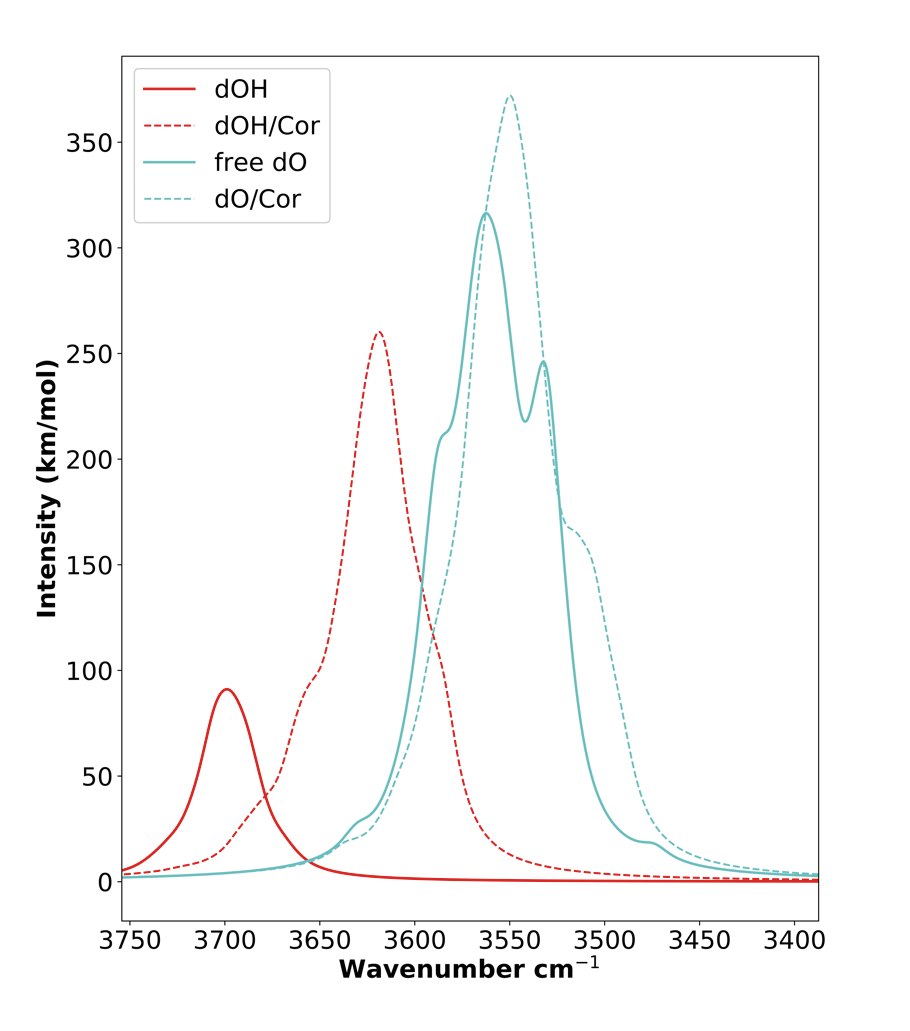}
\caption{Computed IR bands of the dO (blue) and dOH (red) modes of LDA ice before (continuous line) and after (dashed line) adsorption of one Cor molecule (averaged over all configurations). The raw DFTB wavenumbers are scaled by 0.94. Band positions are obtained from the mean position at full width half maximum.  In pure LDA, the dOH and dO  band positions are separated by 149~cm$^{-1}$ (when scaling 0.94 is applied). Upon Cor absorption, the dOH and dO bands are redshifted respectively by 85 (80 with 0.94 factor) cm$^{-1}$ and 8 (8 with 0.94 factor) cm$^{-1}$.}
\label{fig:cor_lda_do}
\end{figure}

An additional complication arises from the fact that the PAH may interact with other ice surface modes (dO, s4),\cite{Buch91,Devlin95} rather than the dOH. 
In our calculations, this was observed most readily when comparing different geometries in the Bz-LDA and Cor-LDA systems and focusing on the dO feature.
Firstly, the position of the dO in pure LDA was well reproduced by the theoretical approach, with the peak of the feature lying 150~cm$^{-1}$ redward of the dOH at $\sim$~3549~cm$^{-1}$ (see Figures~\ref{fig:ben_lda_do} and \ref{fig:cor_lda_do}), in agreement with prior studies.\cite{Devlin95,Noble14a} From our modelling, as was observed for the dOH, the dO modes are typically redshifted upon PAH adsorption. This is illustrated for the cases of Bz and Cor in Figures~\ref{fig:ben_lda_do} and \ref{fig:cor_lda_do}, respectively. 
In most cases, interaction with the aromatic molecule induced a redshift of the dO band (by around 10--20~cm$^{-1}$), as a result of the interaction of the oxygen lone pair with H atoms of the PAH.
%

%%-----------------------------

%%-----------------------------

%%-----------------------------

\begin{figure}[htb!]
\includegraphics[width=0.5\textwidth]{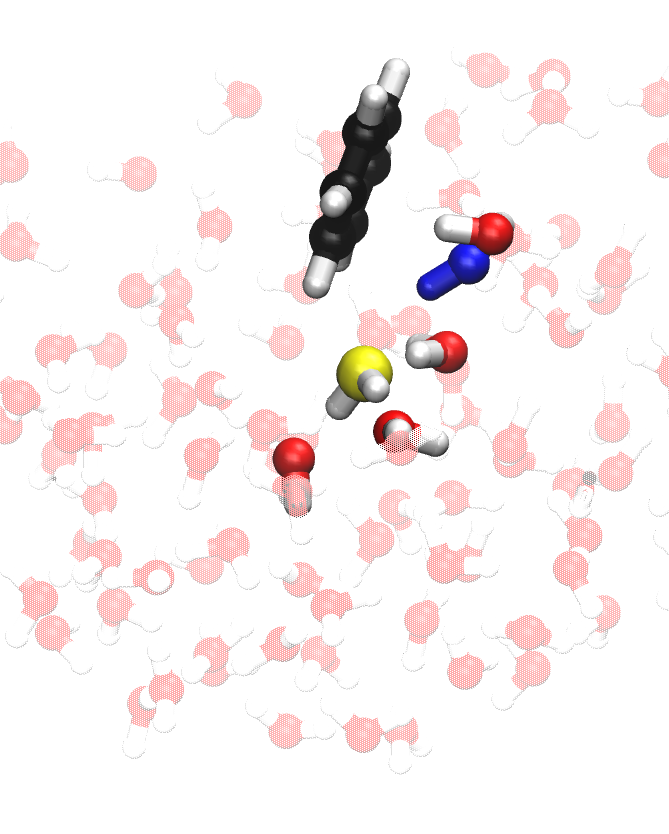}
\caption{Superimposed configurations of LDA with and without a molecule of Bz adsorbed on the surface. 
Each configuration is in its local minimum, and the molecules which are perturbed by the adsorption are magnified and presented in bright colours.
The water molecule which is perturbed the most due to the adsorption of the Bz is represented in blue. The oxygen atom with the highest contribution to the blueshifted dO mode is represented in yellow.}
\label{fig:ben_conf15_dO}
\end{figure}

\begin{figure*}[htb!]
\includegraphics[width=0.43\textwidth]{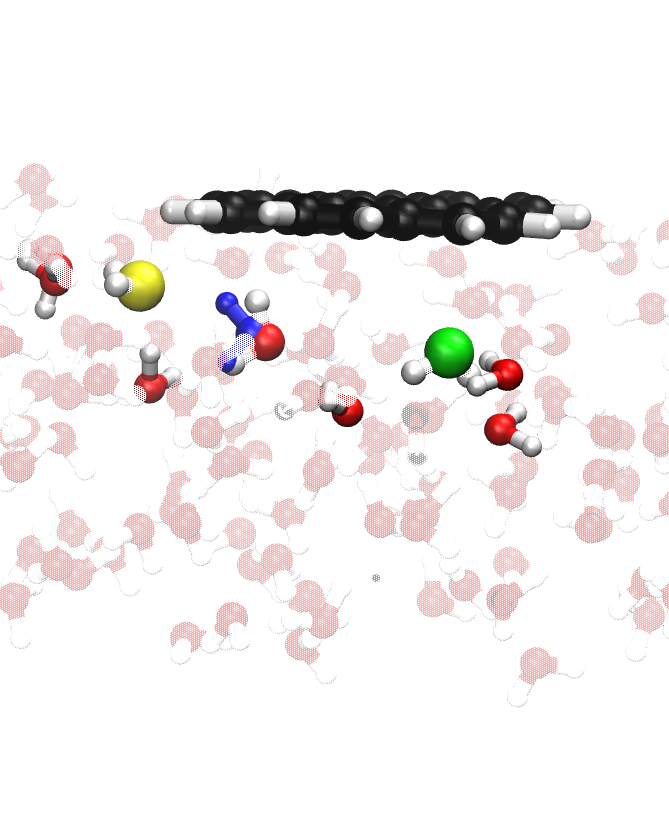}
\includegraphics[width=0.43\textwidth]{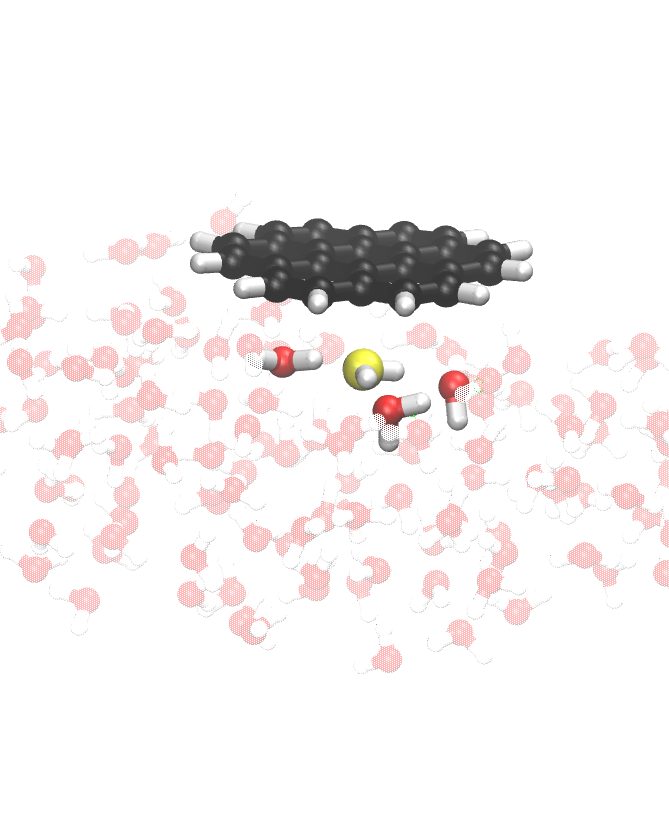} 
\caption{Superimposed configurations of LDA with and without the coronene molecule adsorbed for Configurations Cor$_1$ (left) and Cor$_2$ (right). Each configuration is in its local minimum. Molecules which are perturbed by the adsorption are magnified and presented in bright colours. The water molecule which is perturbed the most due to the adsorption of the coronene is represented in blue. The oxygen atom with the greatest contribution to the blueshifted dO mode is represented in yellow while that in green (Configuration Cor$_1$ only) contributes the most to the redshifted dO mode.}
\label{fig:conf21_dO}
\end{figure*}

In addition to the redshifts, which are the predominant result upon adsorption of an aromatic, blueshifts of the dO modes were also occasionally observed in our modelling. This is illustrated for examples of Bz and Cor adsorption, whose geometries are presented in Figures~\ref{fig:ben_conf15_dO} and \ref{fig:conf21_dO}. 
Some data on the spectral features are reported in Table~\ref{tab:DO_ben_cor_lda}.

For Bz, a blueshift of the dO mode occurs only twice in our sample of 49 configurations, one example of which is illustrated in  Figure~\ref{fig:ben_conf15_dO}. In these two particular configurations, the blueshift is due to the breaking of one hydrogen bond between two water molecules due to the presence of benzene (\textit{i.e.} a new O-H--benzene interaction).

%%-----------------------------

Similarly to the case of Bz, we find that the adsorption of Cor on LDA  can also lead to a blueshift of the dO mode. This again occurs only twice over our 49 configurations, and the two geometries are presented in Figure~\ref{fig:conf21_dO}. 
In the first configuration, Cor$_1$, the blueshift is induced by the breaking of one hydrogen bond between two water molecules due to the presence of coronene (\textit{i.e.} a new O-H--Cor interaction). In this geometry, a redshift is also induced in another dO (labelled in green in Figure~\ref{fig:conf21_dO}). 
In the second configuration, Cor$_2$, the blueshift is due to a repulsive interaction between the oxygen of the closest water molecule and one carbon belonging to the coronene.

The specific geometries highlighted above are observed to give rise to blueshifts on the order of 20--30~cm$^{-1}$. In these rare cases, a carbon atom of the PAH molecule is in interaction with the O atom of the water molecule, suggesting that ices or clusters containing such geometries could exhibit blueshifted dO features. 
It should be reiterated here that, in our calculations, the aromatic molecules almost always favour interaction with a dOH, and the aromatic-dO interactions were noted to occur significantly more rarely than the more favourable aromatic-dOH interactions (two configurations out of 49 each for Bz and Cor).
It is interesting to note, however, that, in these dO interaction configurations, reduction of the bond order of the interacting H$_2$O molecule is also observed \textit{i.e.} intermolecular water-water hydrogen bonds are broken, indicating that adsorption of PAHs, although weak, can perturb the ice surface to the point of disrupting the H-bonding network and thus interact directly with dO adsorption sites despite their large size. This will be discussed in more detail in a forthcoming article (Noble et al. \textit{in prep.}).

\begin{table*}[htb!]
\caption{Wavenumbers (unscaled, in cm$^{-1}$) and intensities  (in km\,mol$^{-1}$) of the dO modes of LDA, isolated and upon adsorption of benzene and coronene, with the most important contributions to the dO blueshift for the four representative configurations detailed in the text.}
\begin{tabular}{|l|l|l||l|l|}
 \hline
 & Wavenumber  & Intensity &  Wavenumber  & Intensity \\ 
 \hline
Config. &\multicolumn{2}{|c||}{ Bz-LDA} & \multicolumn{2}{c|}{LDA }                   \\  \hline

         Bz$_1$      & 3825    & 1252                      & 3778             &     1345      \\ 
         \hline
     Bz$_2$          & 3811             &    650                     & 3776             &     131      \\ \hline\hline
Config. &\multicolumn{2}{|c||}{Cor-LDA} & \multicolumn{2}{c|}{LDA}                   \\  \hline
   Cor$_1$            & 3710             &    1030                     & 3691             &     281      \\ 
               & 3773             &     744                    & 3780             &     669     \\ \hline
     Cor$_2$     & 3779             &     290                & 3759             &    386       \\ \hline
\end{tabular}
\label{tab:DO_ben_cor_lda}
\end{table*}

%

%%-------------------
%%SECTION: CONCLUSION
%%-------------------
\section{Conclusion}

Despite the complexity of varying aromatic-ice interactions arising from bonding preferences,  the key conclusion of this work is that our novel theoretical methodology describing PAH-ice interaction\cite{Michoulier18a,Michoulier18b} is further validated by its ability to accurately model the redshift and widening of the dOH feature of ASW due to the adsorption of PAHs.
Our results highlight the relevance of the configuration sampling approach, wherein averaging the IR harmonic spectra over several tens of configurations reproduces IR band broadening. We clearly confirm that single geometry studies are not sufficient to capture the full range of interactions between an ice surface and a large molecule, and that interaction geometry is a critical issue which must be taken into consideration when studying adsorption and diffusion processes.

This study potentially offers new hope, both for detecting the dOH in icy grain mantles and providing an indicator for the presence of PAHs in space. Although it will not be possible to distinguish individual molecules via this method, we show that it is the dOH feature of the ice that is the most likely ``probe'' of the presence of PAHs, rather than IR bands of PAH molecules themselves, which typically present only small shifts.\cite{Simon17}
When we consider only low mass aromatics (up to about four rings), we are able to distinguish between PAHs of different masses (\textit{i.e.} Anth induces a larger redshift than Bz), but as we increase in PAH size, the redshift values converge. It is likely that the dOH signature -- if observed in interstellar ices -- will be averaged across interaction with PAHs estimated to be in the C$_{50}$-C$_{100}$ range,\cite{Tielens13} and, as such, the dOH will appear at an average redshift value. The width of the band will be large, and thus it will have a low signal to noise, helping to explain why it has not yet been observed.
With the launch of the James Webb Telescope (JWST) imminent, thanks to its high spectral and spatial resolution we will soon have the first opportunity to observe the dOH feature in low-mass star forming regions, affording the chance to answer many outstanding questions on the composition and structure of interstellar ASW.

%%%%%%%%%%%%%%%%%%%%%%%%%%%%%%%%%%%%%%%%%%%%%%%%%%%%%%%%%%%%%%%%%%%%%
\section*{Acknowledgements}

This work has been funded by the French Agence Nationale de la Recherche (ANR) project ``PARCS'' ANR-13-BS08-0005, with support from the French research network EMIE (Edifices Mol\'{e}culaires Isol\'{e}s et Environn\'{e}s, GDR 3533), and the French Programme National ``Physique et Chimie du Milieu Interstellaire'' (PCMI) of the CNRS/INSU with the INC/INP, co-funded by the CEA and the CNES.
E.M. and A.S. acknowledge the computing facility CALMIP at the Paul Sabatier University in Toulouse.

%%%%%%%%%%%%%%%%%%%%%%%%%%%%%%%%%%%%%%%%%%%%%%%%%%%%%%%%%%%%%%%%%%%%%


\begin{thebibliography}{100}

\bibitem{Buch91} Buch, V.; Devlin, J.~P. Spectra of Dangling OH Bonds in Amorphous Ice: Assignment to 2 and 3 Coordinated Surface Molecules. \textit{J. Chem. Phys.} \textbf{1991}, \textit{94}, 4091-4092.
\bibitem{Rowland91} Rowland, B.; Fisher, M.; Devlin, J.~P. Probing Icy Surfaces with the Dangling OH Mode Absorption: Large Ice Clusters and Microporous Amorphous Ice. \textit{J. Chem. Phys.} \textbf{1991}, \textit{95}, 1378-1384.
\bibitem{Rowland95} Rowland, B.; Kadagahur, N.~S.; Devlin, J.~P.; Buch, V.; Feldman, T.; Wojcik, M.~J. Infrared Spectra of Ice Surfaces and Assignment of Surface Localized Modes from Simulated Spectra of Cubic Ice. \textit{J. Chem. Phys.} \textbf{1995}, \textit{102}, 8328-8341.
\bibitem{Devlin95} Devlin, J.~P.; Buch, V. Surface of Ice as Viewed from Combined Spectroscopic and Computer Modeling Studies. \textit{J. Phys. Chem.} \textbf{1995}, \textit{99}, 16534-16548.
\bibitem{Noble14a} Noble, J.~A; Martin, C.; Fraser, H.~J.; Roubin, P.; Coussan, S. Unveiling the Surface Structure of Amorphous Solid Water via Selective Infrared Irradiation of OH Stretching Modes. \textit{J Phys. Chem. Lett.} \textbf{2014}, \textit{5}, 826.
\bibitem{Boogert15} Boogert, A.~C.~A.; Gerakines, P.~A.; Whittet, D.~C.~B. Observations of the Icy Universe. \textit{Annu. Rev. Astron. Astrophys.} \textbf{2015}, \textit{53}, 541.
\bibitem{Keane01} Keane, J.~V.; Boogert, A.~C.~A.; Tielens, A.~G.~G.~M.; Ehrenfreund, P.; Schutte, W.~A. Bands of solid CO$_2$ in the 2--3 $\mu$m spectrum of S 140:IRS1. \textit{Astron. Astrophys.} \textbf{2001}, \textit{375}, L43.
\bibitem{Dulieu10} Dulieu, F.; Amiaud, L.;  Congiu, E.; Fillion, J.-H.; Matar, E.; Momeni, A.; Pirronello, V.; Lemaire, J.~L. Experimental Evidence for Water Formation on Interstellar Dust Grains by Hydrogen and Oxygen Atoms. \textit{Astron. Astrophys.} \textbf{2010}, \textit{512}, A30.
\bibitem{Palumbo06} Palumbo, M.~E. Formation of Compact Solid Water after Ion Irradiation at 15~K. \textit{Astron. Astrophys.} \textbf{2006}, \textit{453}, 903-909.
\bibitem{Accolla11} Accolla, M.; Congiu, E.; Dulieu, F.; Manic\`{o}, F.; Chaabouni, H.; Matar, E.; Mokrane, H.; Lemaire, J.~L.; Pirronello, V. Changes in the Morphology of Interstellar Ice Analogues after Hydrogen Atom Exposure. \textit{Phys. Chem. Chem. Phys.} \textbf{2011}, \textit{13}, 8037-8045.
\bibitem{Noble14b} Noble, J.~A; Martin, C.; Fraser, H.~J.; Roubin, P.; Coussan, S. IR Selective Irradiations of Amorphous Solid Water Dangling Modes:
Irradiation vs Annealing Effects. \textit{J Phys. Chem. C} \textbf{2014}, \textit{118}, 20488.
\bibitem{Dartois15} Dartois, E.; Aug{\'e}, B.; Boduch, P.; Brunetto, R.; Chabot, M.; Domaracka, A.; Ding, J.~J.; Kamalou, O.; Lv, X.~Y.; Rothard, H.; da Silveira, E. F.; Thomas, J. C. Heavy Ion Irradiation of Crystalline Water Ice. Cosmic Ray Amorphisation Cross-section and Sputtering Yield. \textit{Astron. Astrophys.} \textbf{2015}, \textit{576}, A125.
\bibitem{Mejia15} Mej{\'{\i}}a, C.; de Barros, A.~L.~F.; Seperuelo Duarte, E.; da Silveira, E.~F.; Dartois, E.; Domaracka, A.; Rothard, H.; Boduch, P. Compaction of Porous Ices Rich in Water by Swift Heavy Ions. \textit{Icarus} \textbf{2015}, \textit{250}, 222.
\bibitem{Manca01A} Manca, C.; Allouche, A. Quantum Study of the Adsorption of Small Molecules on Ice: The Infrared Frequency of the Surface Hydroxyl Group and the Vibrational Stark Effect. \textit{J. Chem. Phys.}, \textbf{2001}, \textit{114}, 4226-4234.
\bibitem{Martin02} Martin, C.; Manca, C.; Roubin, P. Adsorption of Small Molecules on Amorphous Ice: Volumetric and FT-IR Isotherm Co-measurements Part I. Different Probe Molecules. \textit{Surf. Sci.} \textbf{2002}, \textit{502}, 275-279.
\bibitem{Tielens13} Tielens, A.~G.~G.~M. The Molecular Universe. \textit{Rev. Mod. Phys.} \textbf{2013}, \textit{85}, 1021.
\bibitem{herbig95} Herbig, G.~H. Diffuse Interstellar Bands. \textit{Annu. Rev. Astron. Astrophys.} \textbf{1995}, \textit{33}, 19−73.
\bibitem{Salama11} Salama, F.; Galazutdinov, G. A.; Krelowski, J. et al. Polycyclic Aromatic Hydrocarbons and the Diffuse Interstellar Bands: A Survey. \textit{Astrophys. J.} \textbf{2011}, \textit{728}, 154.
\bibitem{Boogert08} Boogert, A.C.A., Pontoppidan, K.M., Knez, C., et al. The c2d Spitzer Spectroscopic Survey of Ices around Low-Mass Young Stellar Objects. I. H$_{2}$O and the 5-8 {\ensuremath{\mu}}m Bands. \textit{Astrophys. J.} \textbf{2008}, \textit{678}, 985.
\bibitem{Silva94} Silva, S.~C.; Devlin, J.~P. Interaction of Acetylene, Ethylene, and Benzene with Ice Surfaces. \textit{J. Phys. Chem.} \textbf{1994}, \textit{98}, 10847.
\bibitem{Guennoun11c} Guennoun, Z.; Aupetit, C.; Mascetti, J. Photochemistry of Coronene with Water at 10 K: First Tentative Identification by Infrared Spectroscopy of Oxygen Containing Coronene Products. \textit{Phys. Chem. Chem. Phys.} \textbf{2011}, \textit{13}, 7340-7347.
\bibitem{Guennoun11p} Guennoun, Z.; Aupetit, C.; Mascetti, J. Photochemistry of Pyrene with Water at Low Temperature: Study of Atmospherical and Astrochemical Interest. \textit{J. Phys. Chem. A} \textbf{2011}, \textit{115}, 1844-1852.
\bibitem{Michoulier18a} Michoulier, E.; Noble, J.~A.; Simon, A.; Mascetti, J.; Toubin, C. Adsorption of PAHs on Interstellar Ice Viewed by Classical Molecular Dynamics. \textit{Phys. Chem. Chem. Phys.} \textbf{2018}, \textit{20}, 8753.
\bibitem{Michoulier18b} Michoulier, E.; Ben Amor, N.; Rapacioli, M.; Noble, J.~A.; Mascetti, J.; Toubin, C.; Simon, A. Theoretical Determination of Adsorption and Ionisation Energies of Polycyclic Aromatic Hydrocarbons on Water Ice. \textit{Phys. Chem. Chem. Phys.} \textbf{2018}, \textit{20}, 11941.
\bibitem{Simon17} Simon, A.; Noble, J.~A.; Rouaut, G.; Moudens, A.; Aupetit, C.; Iftner, C.; Mascetti, J. Formation of Coronene:water Complexes: FTIR Study in Argon Matrices and Theoretical Characterisation. \textit{Phys. Chem. Chem. Phys.} \textbf{2017}, \textit{19}, 8516.
\bibitem{Bouwman11} Bouwman, J.; Mattioda, A.~L.; Linnartz, H.; Allamandola, L.~J. Photochemistry of Polycyclic Aromatic Hydrocarbons in Cosmic Water Ice I. Mid-IR Spectroscopy and Photoproducts. \textit{Astronom. Astrophys.} \textbf{2011}, \textit{525}, A93.
\bibitem{Cook15} Cook, A.~M.; Ricca, A.; Mattioda, A.~L., et al. Photochemistry of Polycyclic Aromatic Hydrocarbons in Cosmic Water Ice: The Role of PAH Ionization and Concentration.\textit{Astrophys. J.} \textbf{2015}, \textit{799}, 14.
\bibitem{deBarros17} de Barros, A.~L.~F.; Mattioda, A.~L.; Ricca, A.; Cruz-Diaz, G.~A.; Allamandola, L.~J. Photochemistry of Coronene in Cosmic Water Ice Analogs at Different Concentrations. \textit{Astrophys. J.} \textbf{2017}, \textit{848}, 112.
\bibitem{Yamada83} Yamada, H.; Saheki, M. Infrared ATR Measurements at Low Temperatures. An Apparatus and the Absorption Intensities of Crystalline Benzene and Carbon Disulfide. \textit{Bull. Chem. Soc. Jpn.} \textbf{1983}, \textit{56}, 35. 
\bibitem{Elstner98} Elstner, M.; Porezag, D.; Jungnickel G.; Elsner, J.; Haugk, M.; Frauenheim, Th.; Suhai, S.; G. Seifert, G. Self-Consistent-Charge Density- Functional Tight-Binding Method for Simulations of Complex Material Properties. \textit{Phys. Rev. B} \textbf{1998}, \textit{58}, 7260. 
\bibitem{Li_CM3} Li, J.; Zhu, T.; Cramer, C.~J.; Truhlar, D.~G. New Class IV Charge Model for Extracting Accurate Partial Charges from Wave Functions. \textit{J. Phys. Chem. A} \textbf{1998}, \textit{102}, 1820-1831.
\bibitem{DFTB_CM3} Rapacioli, M.; Spiegelman, F.; Talbi, D.; Mineva, T.; Goursot, A.; Heine, T.; Seifert, G. Correction for Dispersion and Coulombic Interactions in Molecular Clusters with Density Functional Derived Methods: Application to Polycyclic Aromatic Hydrocarbon Clusters. \textit{J. Chem. Phys.} \textbf{2009}, \textit{130}, 244304--10.
\bibitem{joalland2010} Joalland, B.; Rapacioli, M.; Simon, A.; Joblin, C.; Marsden, C.~J.; Spiegelman, F. Molecular Dynamics Simulations of Anharmonic Infrared Spectra of [SiPAH] pi-Complexes. \textit{J. Phys. Chem. A} \textbf{2014}, \textit{114}, 5846-5854.
\bibitem{SimonPCCP2012} Simon, A.; Rapacioli, M.; Mascetti, J.; Spiegelman, F. Vibrational Spectroscopy and Molecular Dynamics of Water Monomers and Dimers Adsorbed on Polycyclic Aromatic Hydrocarbons. \textit{Phys. Chem. Chem. Phys.} \textbf{2012}, \textit{14}, 6771-6786.
\bibitem{SimonJCP2013} Simon, A.; Spiegelman, F. Water Clusters Adsorbed on Polycyclic Aromatic Hydrocarbons: Energetics and Conformational Dynamics. \textit{J. Chem. Phys.} \textbf{2013}, \textit{138}, 194309.
\bibitem{Simon19} Simon, A.; Rapacioli, M.; Michoulier, E.; Zheng, L.; Korchagina, K.; Cuny, J. Contribution of the Density-Functional based Tight-Binding Scheme to the Description of Water Clusters: Methods, Applications and Extension to Bulk Systems \textit{Mol. Sim.} \textbf{2019}, \textit{45}, 249-268.
\bibitem{Gruenloh97} Gruenloh, C.~J.; Carney, J.~R.; Arrington, C.~A.; Zwier, T.~S.; Fredericks, S.~Y.; Jordan, K.~D. Infrared Spectrum of a Molecular Ice Cube: The S$_4$ and D$_{2d}$ Water Octamers in Benzene-(Water)$_8$. \textit{Science} \textbf{1997}, \textit{276}, 1678. 
\bibitem{Gruenloh98} Gruenloh, C.~J.; Carney, J.~R.; Hagemeister, F.~C.; Arrington, C.~A.; Zwier, T.~S.; Fredericks, S.~Y.;  Wood, J.~T.; Jordan, K.~D. Resonant Ion-dip Infrared Spectroscopy of the S$_4$ and D$_{2d}$ water octamers in benzene-(water)$_8$ and benzene$_2$-(water)$_8$. \textit{J. Chem. Phys.} \textbf{1998}, \textit{109}, 6601.
\bibitem{Miliordos16} Miliordos, E.; Apra E.; Xantheas, S.~S. A New, Dispersion-Driven Intermolecular Arrangement for the Benzene-Water Octamer Complex: Isomers and Analysis of their Vibrational Spectra. \textit{J. Chem. Theory Comput.} \textbf{2016}, \textit{12}, 4004.
\bibitem{Zwier96} Fredericks, S.~Y.; Jordan, K.~D.; Zwier, T.~S. 
Theoretical Characterization of the Structures and Vibrational Spectra of Benzene-(H$_2$O)$_n$ (n=1-3) Clusters. \textit{J. Phys. Chem.} \textbf{1996}, \textit{100}, 7810.
\bibitem{Oliveira2015} Oliveira, L.; Fernando L.; Cuny, J.; Moriniere, M.; Dontot, L.; Simon, A.; Spiegelman, F.; Rapacioli, M. Phase Changes of the Water Hexamer and Octamer in the Gas Phase and Adsorbed on Polycyclic Aromatic Hydrocarbons. \textit{Phys. Chem. Chem. Phys.} \textbf{2015}, \textit{17}, 17079-17089.
\bibitem{Hujo11} Hujo, W.; Gaus, M.; Schultze, M.; Kubar, T.; Grunenberg, J.; Elstner, M.; Bauerecker, S. Effect of Nitrogen Adsorption on the Mid-Infrared Spectrum of Water Clusters. \textit{J. Phys. Chem. A} \textbf{2011}, \textit{115}, 6218-6225.
\bibitem{Pezzella2018} Pezzella, M.; Unke, O.T.; Meuwly M. Molecular Oxygen Formation in Interstellar Ices Does Not Require Tunneling. \textit{J Phys. Chem. Lett.} \textbf{2018}, \textit{8}, 1822-1826.
\bibitem{Dawes18} Dawes, A.; Pascual, N.; Mason, N.~J.; G\"{a}rtner, S.; Hoffmann, S.~V.; Jones, N.~C. Probing the Interaction between Solid Benzene and Water using Vacuum Ultraviolet and Infrared Spectroscopy. \textit{Phys. Chem. Chem. Phys.} \textbf{2018}, \textit{20}, 15273.
\bibitem{Pribble94} Pribble, R.~N.; Zwier, T.~S.  Size-Specific Infrared Spectra of Benzene-(H$_2$O)$_n$ Clusters (n = 1 through 7): Evidence for Noncyclic (H$_2$O)$_n$ Structures. \textit{Science} \textbf{1994}, \textit{265}, 75-79.
\bibitem{Gruenloh00} Gruenloh, C.~J.; Carney, J.~R.; Hagemeister, F.~C.;  Zwier, T.~S.; Wood, J. T.; Jordan, K.~D.  Resonant Ion-dip Infrared Spectroscopy of bzw9: Expanding the Cube.  \textit{J. Chem. Phys.} \textbf{2000}, \textit{113}, 2290.
\bibitem{Noble17coro} Noble, J.~A.; Jouvet, C.; Aupetit, C.; Moudens, A.; Mascetti, J. Efficient Photochemistry of Coronene:water Complexes. \textit{Astron. Astrophys.} \textbf{2017}, \textit{599}, A124.

\end{thebibliography}
\end{document}